\documentclass[12pt,fleqn]{article}

\usepackage[T1]{fontenc}
\usepackage[ansinew]{inputenc}
\usepackage{amsmath}
\usepackage{indentfirst}
\usepackage{natbib}
\usepackage{graphics}
\usepackage{enumerate}

\begin{document}

\title{\bf THEORY OF LOCALIZATION IN A DISORDERED RING IN A NON-HERMITIAN FIELD}
\author{J.Heinrichs*\\Institut de Physique B5, University of Liege,\\ Sart-Tilman, B-4000 Liege, Belgium}
\date{}
\maketitle

\begin{abstract}
\thispagestyle{empty}
\noindent
We present an analytical study of the Lyapunov exponent $\lambda$ for electronic states in a disordered tight-binding ring,
with $N$ sites of spacing $c$, in the presence of a non-hermitian field component $h$, for weak disorder.  This system has
been proposed by Hatano and Nelson (HN) as a model for the depinning of flux lines bound to columnar defects in a
superconducting cylindrical shell in a transverse magnetic field proportional to $h$.  We calculate the Lyapunov exponent
to second order in the disorder in the domain of complex energies in an ordered ring in a non-hermitian field.  The result exhibits an important 
symmetry property, $\lambda (-h)=-\lambda (h)$, which reflects the directionality of the non-hermitian localization.  By combining the results for $\lambda$ with a
previous calculation of the complex eigenvalues for weak disorder, we show that the localization length of the complex eigenenergy states is generally larger than
the circumference ($Nc$) of the ring.  This supports the initial suggestion of HN that these eigenstates are delocalized.  Our perturbative treatment of the
disorder when applied for an open chain embedded in an infinite non-disordered chain recovers the Thouless formula for the zero-field localization length.\par
\vspace{1cm}
PACS numbers: 72.15 Rn, 74.60 GE, 71.30.+h\par

*e-mail: J.Heinrichs@ulg.ac.be
\end{abstract}

\newpage
\pagestyle{plain}
\setcounter{page}{1}    
\section{INTRODUCTION}

It has been known since the classic papers of Mott and Twose\citep{Mott} and of Borland\citep{Borland}
that disorder localizes electronic states in an open linear chain.  This observation was reinforced about two
decades later, by detailed analytical calculations of the inverse localization length (ILL) for weakly
disordered tight-binding open chains [3-7], for energies in the band of the ordered system. 

Recently, renewed interest in one-dimensional
localization has arisen in two important contexts involving closed circular chains (rings) rather than open linear chains. 
The first one is the problem of the universal persistent current induced in thin mesoscopic metallic rings threaded by an
Aharonov-Bohm~flux, which has been studied both theoretically [8-10] and experimentally
\citep{Levy}.  The other one relates to flux lines pinned to random columnar defects in a superconducting
cylindrical shell\citep{Hatano}.  This system can be modelled by a non hermitian quantum hamiltonian for an electron
in a disordered ring in the presence of a constant imaginary vector potential (IVP) (proportional to the
external magnetic field acting transversely to the flux lines in the superconductor\citep{Hatano}).  The
localization-delocalization transition (corresponding to a pinning-depinning transition for the flux lines)
predicted by Hatano and Nelson\citep{Hatano} at a threshold vector potential, has generated considerable
interest from an early stage\citep{Efetov,Brouwer} till recently\citep{Goldsheid,Mudry}.  Localization in the
presence of an imaginary vector potential is referred to specifically as directed localization\citep{Efetov}
because, unlike in a real vector potential, the symmetry between right and left moving particles is broken
while time reversal symmetry is not. 

The previous analyses of the localization-delocalization transition in the non-hermitian hamiltonian of Hatano and
Nelson[12-15] are mainly based on the study of the eigenvalues which may be complex.  
Indeed it has been argued that real eigenvalues correspond to localized states while complex eigenvalues
correspond to extended states\citep{Hatano}.  In particular, both the analyses of the directed-localization
problem and those of the effect of disorder on persistent currents [9,10] use the fact that in zero field the electronic states in a one-dimensional
disordered system are localized and correspond to real eigenvalues.\par

The purpose of this paper is to discuss the theory of localization and of directed localization by
means of a first principles analytical study of the Lyapunov exponent $\lambda$ of the electronic states in a
weakly disordered tight-binding ring in a non-hermitian field.  The ILL in the non-hermitian field, which we
shall refer to more specifically as the inverse directed localization length (IDLL), $1/\xi$, is related to
$\lambda$ by $1/\xi=|\lambda|$.\par

\quad The Schršdinger equation for the hamiltonian of a disordered one-orbital tight-binding ring of circumference $Nc$
(with $N$ the number of sites and $c$ the lattice spacing) in an IVP reduces to the set of equations

\begin{equation}\label{eq.1}
e^{hc}\varphi_{n+1}+e^{-hc}\varphi_{n-1}+\varepsilon_n\varphi_n =E\varphi_n\;,\;n=2,3,\ldots N-1,
\end{equation}
\begin{equation}\label{eq.2}
e^{hc}\varphi_2+e^{-hc}\varphi_N+\varepsilon_1\varphi_1=E\varphi_1,
\end{equation}
\begin{equation}\label{eq.3}
e^{hc}\varphi_1+e^{-hc}\varphi_{N-1}+\varepsilon_N\varphi_N=E\varphi_N.
\end{equation}

\noindent Here $\varphi_n$ is the amplitude of a wavefunction at site $n,\;E$  and the parameters $\varepsilon_n$  are the
energy and the random site energies in units of minus the nearest-neighbour hopping parameter, respectively. The hamiltonian
in (\ref{eq.1}-\ref{eq.3}) is non hermitian because of the real field parameter $h$, corresponding to an
IVP\citep{Hatano}.\par

In an interesting recent paper, Brouwer~{\it et al.} presented an analytic study of the complex eigenvalues of the non-hermitian tight-binding ring for weak
disorder [14].  This study is complemented here by a corresponding analytic study of the IDLL for complex energy states for weak disorder.  Our treatment
is valid for sufficiently strong fields such that the disorder may be treated perturbatively.  The localization length of complex energy states as defined in a
ring of infinite circumference \newline ($Nc=\infty$) is generally finite.  This is similar to the decay length $\xi$ of the transmission coefficient introduced by
Brouwer~{\it et al.} in their analysis of complex eigenvalues which is also found to be finite and given by $\xi=(2|h|)^{-1}$\citep{Brouwer}.  However, when
applying these results to a finite ring one must consider the restriction on the validity of the perturbation treatment of the disorder which is imposed by the
discreteness of the energy spectrum of (\ref{eq.1}-\ref{eq.3}).  Specifically, the corrections to the zeroth order complex energy levels due to the disorder
must be small compared to the level spacing of the ring.  Using a previous calculation \citep{JH2} of the eigenvalues of (\ref{eq.1}-\ref{eq.3}) for weak
disorder, this condition allows us to show that the localization lengths of the complex eigenenergy states are generally larger than the circumference of the
ring at the considered field strengths.  These results explicitly support the suggestion of HN \citep{Hatano} that the complex energy states in a finite ring are
delocalized.\par

The paper is organized as follows. In Sect.II we present the detailed solutions for the amplitudes $\varphi_n$ in
(\ref{eq.1}-\ref{eq.3}) for weak disorder, at arbitrary energy E both in the domain of the complex spectrum of the ordered
system, and in a range of real energies.  From these solutions we obtain the explicit expressions for the Lyapunov exponent
for the eigenstates of the ring, for $h\neq 0$, to second order in the random site energies.  In Sect.III we first apply a similar
perturbation analysis for weak disorder to rederive the Thouless~formula for the zero field ILL for the case of an open
chain.  This further supports the validity of our calculation for the case of a ring.  Next we discuss the properties of our
general expression for the IDLL and its application to complex eigenenergy states in finite rings.  In Sect.~IV we 
conclude with some final remarks.

\section{THE NON-HERMITIAN LYAPUNOV \\ EXPONENT FOR WEAK DISORDER}

In solving (\ref{eq.1}-\ref{eq.3}) recursively, initiating the recursion at site $n=1$ with an arbitrary $\varphi_1$,
it is convenient  to define the amplitude ratios $R_n=\varphi_n/\varphi_{n-1}\;,\;n=2,3,\ldots
N,\;\text{and}\;Q_1=\varphi_1/\varphi_N$ obeying the equations

\begin{equation}\label{eq.4}
e^{hc}R_{n+1}+e^{-hc}R^{-1}_n=E-\varepsilon_n\;,\;,n=2,3,\ldots N-1,
\end{equation}
\begin{equation}\label{eq.5}
e^{hc}R_2+e^{-hc}Q^{-1}_1=E-\varepsilon_1,
\end{equation}
\begin{equation}\label{eq.6}
e^{hc}Q_1+e^{-hc}R^{-1}_N=E-\varepsilon_N,
\end{equation}

\noindent which determine the quantities $R_n\;,\;n=2,3,\ldots N$, and $Q_1$ as function of an arbitrary energy $E$, if one
disregards the secular equation giving the eigenvalues, which may be written in the form

\begin{equation}\label{eq.7}
e^{hc}R_2(E)+e^{-hc}\prod^N_{m=2}\;R_m(E)=E-\varepsilon_1,
\end{equation}

\noindent using (\ref{eq.2}) and the general definition

\begin{equation}\label{eq.8}
\varphi_n=\prod^n_{m=2}\;R_m.\varphi_1\;,\;n=2,3,\ldots N\quad .
\end{equation}

\noindent The secular equation (\ref{eq.7}) will be automatically verified by the quantities $R_n(E)$ above whenever $E$ coincides 
with one of the eigenvalues $E=E_\alpha$.

The reason for studying the exponential variation of the quantities $R_m$ and of the corresponding amplitudes
$\varphi_n$ as a function of a generic energy $E$, at sites $n$ further and further away from the initial site, is that it
permits to obtain the Lyapunov exponent of the electronic eigenstates of the ring on the basis of the Mott-Twose-Borland
conjecture\citep{Mott,Borland,Crisanti}.  Indeed, Mott, Twose and Borland argue that when $E$ is
close to an eigenvalue the exponential rate of variation of the amplitudes $\varphi_n$ at far away sites $n$ tends to the
(largest) exponential rate of localization of the corresponding eigenstate of the ring about a
fixed localization centre.  The Lyapunov exponent which characterizes the exponential growth (or decay) of $\varphi_n$ at
distant sites ($nc\rightarrow\infty$), at real or complex energies $E$, is given by\citep{1}

\begin{align}\label{eq.9}
\lambda &=\lim_{(n\leq N)\rightarrow\infty}\frac{1}{nc}\ln |\varphi_n|,\nonumber\\
&=\lim_{n\rightarrow\infty}\frac{1}{nc}\sum^n_{p=2}\ln |R_p|\quad .
\end{align}

\noindent and since $\lambda$ self-averages to a central limit value\citep{Furstenberg}  we further have

\begin{equation}\label{eq.10}
\lambda=\lim_{n\rightarrow\infty}\frac{1}{nc}\sum^n_{p=2}\langle \ln |R_p|\rangle\quad .
\end{equation}

\noindent The Lyapunov exponent may generally take positive- as well as negative values.  Since all sites of a ring are
equivalent by definition, there is no fundamental distinction between states corresponding to different signs of the Lyapunov
exponents.  In contrast, for an open chain, a positive Lyapunov exponent is associated with states which are growing as one is
moving away from an edge site $n=1$ while a negative Lyapunov exponent corresponds to a state which decays with increasing
distance from the edge.  The former state is generally referred to as a localized state while the latter might be termed an
antilocalized state\citep{Heinrichs} to emphasize the fact its amplitude is smallest deep in the interior of the chain.\par

Assuming the disorder to be weak, we now analyze the perturbative solutions of (4-6) to second order in the site energies
$\varepsilon_i$.  Since we are interested in the study of the localization length, we only consider solutions as a function of
a generic complex energy $E$, thus ignoring the consistency-or eigenvalue equation (\ref{eq.7}).  In the absence of disorder
the solutions of the non-hermitian system (4-6) are of the form

\begin{equation}\label{eq.11}
R^0_n=Q^0_1=e^{i\;q}\;,\;n=2,3\ldots N,
\end{equation}

\noindent and correspond to a generic complex energy 

\begin{equation}\label{eq.12}
E=2\cos (q-ihc)
\end{equation}

\noindent expressed in terms of an arbitrary real wavenumber $q$ in units of the lattice para\-meter.  The equation (\ref{eq.12})
defines energies lying on an ellipse,

\begin{equation}\label{eq.13}
\frac{(ReE)^2}{\cosh^2hc}+\frac{(ImE)^2}{\sinh^2hc}=4,
\end{equation}

\noindent whose semi-axes $2\cosh hc$ and $2\sinh |h|c$ give the half-width of the energy band of edges $\pm 2\cosh hc$ on
the real axis and of the corresponding band with edges $\pm\sinh |h|c$ on the imaginary energy axis, respectively.  The
meaning of the energies (\ref{eq.12}) is that they correspond to the domain of extended Bloch state eigenvalues in the absence
of disorder, which are given by the special values for $q=2\pi k/N\;,$ \linebreak $\;k=0,1,2,\ldots$ as a consequence of single-valuedness
of the wave functions on the ring, expressed by (\ref{eq.7}).  On the other hand, the meaning of $ImE$ has been discussed
by Hatano and Nelson in the context of depinned flux lines\citep{Hatano}.  In fact, Hatano and Nelson argue
convincingly that complex energies are a necessary feature of the delocalized states which may appear in the depinning
transition.\par 

The corrections in $R_n$ and $Q_1$ at first and the second orders in the energies $\varepsilon_i,i=1,2,\ldots N$, are
determined by inserting the expansions \linebreak
$R_n=e^{i\;q}+R^{(1)}_n+R^{(2)}_n\;,\;n=2,3,\ldots N\;,\;Q_1=e^{i\;q}+Q^{(1)}_1+Q^{(2)}_1$,
involving linear and quadratic terms, in the equations (4-6) and
identifying terms of the same order in the order by order expansions of them.  This leads to the following sets of equations at first and second orders,
respectively:

\begin{subequations}
\renewcommand{\theequation}
{\theparentequation.\alph{equation}}\label{eq.14}
\begin{align}
R_{n+1}^{(1)}-\tilde a\;R^{(1)}_n &=-e^{-hc}\varepsilon_n\;,\;n=2,3,\ldots N-1,\\
R^{(1)}_2-\tilde a^2\;R^{(1)}_N &=-e^{-hc}(\tilde a\;\varepsilon_N+\varepsilon_1),
\end{align}
\end{subequations}
\begin{subequations}\label{eq.15}
\renewcommand{\theequation}
{\theparentequation.\alph{equation}}
\begin{align}
R_{n+1}^{(2)}-\tilde a\;R^{(2)}_n &=-\tilde a\;e^{-iq} R_n^{(1)2}\;,\;n=2,3,\ldots N-1,\\
R^{(2)}_2-\tilde a^2\;R^{(2)}_N &=-\tilde a^2(1+\tilde a)\;e^{-iq}R_N^{(1)2}
+\tilde a\;\sqrt{\tilde a}\;\varepsilon_N(2\tilde a\;R_N^{(1)}-e^{-hc}\varepsilon_N).
\end{align}
\end{subequations}
 
\noindent Note that the quantities $Q_1^{(1)}$ and $Q_1^{(2)}$ have been eliminated since they do not enter in the
determination of the wavefunction amplitudes  (\ref{eq.8}).  In the above equations we have defined

\begin{equation}\label{eq.16}
\tilde a =\exp(-2i\tilde q)\;,\;\tilde q=q-ihc,
\end{equation}

\noindent where $\tilde q$ in a wavenumber with an imaginary part arising from the non-hermitian field.

By solving successively the coupled first and second order difference equations (14.a,b) and (15.a,b) for the
quantities
$R^{(1)}_n$ and $ R^{(2)}_n $ we get:

\begin{subequations}\label{eq.17}
\renewcommand{\theequation}
{\theparentequation.\alph{equation}}
\begin{align}
R^{(1)}_n &=\tilde a^{n-2} R^{(1)}_2-e^{-hc}\sum^{n-1}_{m=2}\;\tilde a^{n-m-1}\;\varepsilon_m\;,\;n=3,4,\ldots N,\\
R^{(1)}_2 &=\frac{e^{-hc}}{\tilde a^N-1}\biggl(\varepsilon_1+\tilde a\varepsilon_N
+\tilde a\sum^{N-1}_{m=2}\;\tilde a^{N-m}\varepsilon_m\biggr),
\end{align}
\end{subequations}
\begin{subequations}\label{eq.18}
\renewcommand{\theequation}
{\theparentequation.\alph{equation}}
\begin{align}
R^{(2)}_n &=\tilde a^{n-2} R^{(2)}_2-e^{hc}\sqrt{\tilde a}\sum^{n-1}_{m=2}\;\tilde a^{n-m}\biggl(R^{(1)}_m\biggr)^2\;,\;n=3,4,\ldots N,\\
R^{(2)}_2 &=\frac{e^{hc}\sqrt{\tilde a}}{\tilde a^N-1}
\biggl[\frac{1}{\tilde a}\biggl(e^{-hc}\varepsilon_1+R^{(1)}_2\biggr)^2
+\sum^N_{m=2}\;\tilde a^{N-m+2}\biggl(R^{(1)}_m\biggr)^2\biggr].
\end{align}
\end{subequations}

\noindent It should be noted that the expression for $ R^{(1)}_2$ and $R^{(2)}_2$, equations (17.b) and (18.b), are invalid at
zero field, where they diverge at the energies $E=2\cos q$, \linebreak $q=\frac{2\pi.k}{N}\;,\;k=0,1,2,\ldots$ corresponding to the eigenvalues of the
ordered ring for
$h=0$.  In fact, as is well known, perturbation theory for eigenstate energies in a weakly disordered ring also diverges for $h=0$\citep{JH2}.  This strongly
suggests that the Lyapunov exponents of the zero field eigenstates in the ring may not be discussed by perturbation theory.\par

The next step is then to expand $\ln (Re\;R_p)$ in (\ref{eq.10}) up to second order terms\citep{2},
\begin{align}\label{eq.19}
\ln |R_p| &=\frac{1}{2}(e^{-iq}R_p^{(1)}+ c.c.)\nonumber\\
&+\frac{1}{2}\biggl[e^{-iq}\biggl(R^{(2)}_p-\frac{e^{-iq}}{2}R_p^{(1)2}\biggr)+c.c.\biggr]
\;,\;p=2,3,\ldots N,
\end{align}

\noindent and to obtain the averages of the  quantities on the r.h.s. using the explicit expressions (17.a), (17.b) and 
(18.a), (18.b).  Here we assume the $\varepsilon_i$Õs to be identically distributed, independent gaussian variables
with mean zero and a correlation $\langle\varepsilon_i\varepsilon_j\rangle=\varepsilon^2_0\;\delta_{i,j}$, where the choice
$\varepsilon^2_0=W^2/12$ would correspond to using the familiar rectangular distribution of width $W$.  The determination of
the non-zero averages proceeds by first finding the averages $\langle R^{(1)2}_2\rangle$ and
$\langle R^{(1)2}_p\rangle\;,\;p\neq 2$, from (17.a) and (17.b), which are then inserted in (18.a) and (18.b) for
obtaining the averages $\langle R^{(2)}_2\rangle$ and $\langle R^{(2)}_p\rangle\;,\;p\neq 2$.  The sums over lattice sites in
these various averages are geometric progressions which are easily handled.  The final form of the averages of interest are
remarkably simple since we obtain

\begin{subequations}\label{eq.20}
\renewcommand{\theequation}
{\theparentequation.\alph{equation}}
\begin{align}
\langle R^{(1)2}_p\rangle &=\langle R^{(1)2}_2\rangle=\langle R^{(1)2}_N\rangle
=\frac{\varepsilon^2_0\;e^{-2hc}}{(1-\tilde a^2)}\biggl(\frac{1+\tilde a^N}{1-\tilde a^N}\biggr),\\
\langle R^{(2)}_p\rangle &=\langle R^{(2)}_2\rangle=\langle R^{(2)}_N\rangle=
\frac{\varepsilon_0^2\;e^{-hc}\tilde a\sqrt{\tilde a}}{(1-\tilde a)(1-\tilde a^2)}
\biggl(\frac{1+\tilde a^N }{1-\tilde a^N }\biggr)\;,\;p=3,4,\ldots N-1.
\end{align}
\end{subequations}

\noindent This reflects the independence of these quantities of the initial site(s) for the recursive solution of the tight-binding equations, as expected.  By
substituting these results in (\ref{eq.19}) and (\ref{eq.10}) we obtain our final expression for Lyapunov exponent
$(N\rightarrow\infty)$ to second order in the disorder, 

\begin{equation}\label{eq.21}
\lambda=-\frac{\varepsilon^2_0}{4}\frac{\tilde a}{(1-\tilde a)^2}
\biggl(\frac{1+\tilde a^N}{1-\tilde a^N}\biggr)+c.c.,
\end{equation}

\noindent where we have used the definition (\ref{eq.16}).  The equation (\ref{eq.21}) is valid for complex energies
corresponding to  the energy band (\ref{eq.13}) of the pure ring in a non-hermitian field $h\neq 0$.\par

The above analysis may be extended in principle to the description of the IDLL for real eigenvalue states.  It suffices to 
choose the parameter $q$ in (\ref{eq.11}) to be complex with the forms

\begin{subequations}\label{eq.22}
\renewcommand{\theequation}
{\theparentequation.\alph{equation}}
\begin{alignat}{2}
q&=i\kappa &\qquad ,\\
q&=\pi+i\kappa &\qquad ,
\end{alignat}
\end{subequations}

\noindent where $\kappa$ is real.  Indeed the values (22.a,b) define generic real energies 

\begin{equation}\label{eq.23}
E=\pm2\cosh\;(\kappa-hc)\quad ,
\end{equation}

\noindent which together with (\ref{eq.11}) and (22.a,b) are solutions of (4-6) for the pure system.  Eigenvalues of
the  non-disordered ring obtained by inserting (\ref{eq.11}), (22.a,b) and (\ref{eq.23}) in (\ref{eq.7}) correspond to
$\kappa=0$ i.e. to the real band edges $\pm 2\cosh hc$.  On the other hand, the perturbation of eigenstate energies for weak disorder does not lead to real
eigenvalues apart from the edges ($\pm 2\cosh\;hc$) of the $Re\;E$-domain of the complex eigenvalues [14,17].  Therefore the perturbation study of the IDLL
for real energies is of no interest.

\section{DISCUSSION OF RESULTS}

\subsection{Validity of perturbation theory}
\quad In this subsection we demonstrate the validity of the perturbation expansion of Sect.II by showing that when applied to
the case of an open disordered chain it recovers the well-known Thouless~formula for the inverse localization length in the
absence of an IVP.  Here an open disordered chain is defined as a segment of length $Nc$ which is connected at both ends to
semi-infinite non-disordered discrete chains.  The presence of semi-infinite non-disordered chains is indeed required for
the existence of a transmission coefficient $T$ related to the inverse localization length by $T\approx e^{-\lambda
Nc}\;\text{for}\; N\rightarrow\infty$\citep{Ishii}.  This implies that for an open chain the solution of equation
(\ref{eq.4}) in the absence of disorder is still given by (for $h=0$)

\begin{equation}\label{eq.26}
R^{(0)}_n=e^{iq}\;,\;E=2\cos q\;,\;n=1,2,\ldots N.
\end{equation}

\noindent In order to study the Lyapunov exponent for a large open chain for weak disorder it is then sufficient to solve the
perturbation equations (14.a) and (15.a) alone in terms of arbitrary finite $\varphi_1$ and $R^{(1)}_2, R^{(2)}_2$ in the
expansion

\begin{equation}\label{eq.27}
R_2=e^{iq}+R_2^{(1)}+ R_2^{(2)}+\ldots\qquad .
\end{equation}

\noindent The relevant averages in (\ref{eq.10}), using (\ref{eq.19}) are obtained from (17.a) and (18.a) for $h=0$ and are given
by (with $a=e^{-2iq}$)

\begin{equation}\label{eq.28}
\langle R_p^{(1)2}\rangle= a^{2(p-2)} R_2^{(1)2}+\varepsilon_0^2\;e^{-2h}
\biggl(\frac{1- a^{2(p-2)}}{1- a^2}\biggr),
\end{equation}
\begin{multline}\label{eq.29}
\langle R_p^{(2)}\rangle= a^{2(p-2)}(R_2^{(2)}-e^{hc}\sqrt{ a} R_2^{(1)2})\\
+\varepsilon^2_0\;e^{-hc}\;\frac{\sqrt{ a}\; a^{p-1}}{(1- a^2)(1- a)}
[1+ a^{-1}- a^{p-3}- a^{-p+2}].
\end{multline}

\noindent By inserting these results in (\ref{eq.19}) and noting that only terms independent of $p$ contribute to the Lyapunov 
(\ref{eq.10}) we finally obtain, using \eqref{eq.26},

\begin{equation}\label{eq.30}
\lambda=\frac{1}{\xi}=\frac{\varepsilon^2_0}{2}\frac{1}{4-E^2}\qquad ,
\end{equation}

\noindent which is the well-known Thouless formula\citep{Thouless,Luck}, for
the ILL showing that all eigenstates with eigenvalues lying in the band
\eqref{eq.26} are exponentially localized (with the usual proviso that the Mott-Twose-Borland argument about the connection
between eigenstates and exponentially growing solutions from both ends of a long open chain, at an energy $E$, is valid).

\subsection{Directed localization on a ring}
In the presence of an IVP, the rates of hopping from a given site to its nearest neighbour to the left- and to the
right hand sides on the ring are different.  This asymmetry of the hopping rates is reflected in the antisymmetry of the
Lyapunov exponent of the eigenstates under reversal of the non-hermitian field $h$,

\begin{equation}\label{eq.31}
\lambda(-h)=-\lambda(h),
\end{equation}

\noindent which is demonstrated explicitly by the equation (\ref{eq.21}), to lowest non-vanishing order in the gaussian
disorder.  To our knowledge this important antisymmetry property has not been discussed before.  The physical interpretation of \eqref{eq.31} is simple.  We might
expect the Lyapunov exponent in \eqref{eq.21} to describe a symmetric exponentially localized state: this means that since $\lambda (h)$ determines the exponential
growth (if $\lambda (h)>0$) of an initial amplitude $\varphi_1$ with increasing $n=1,2,3,\ldots$ then the corresponding exponent describing the
growth of the localized state with decreasing $n=N,\;N-1,\;N-2,\ldots$ should be $-\lambda (h)$.  The fact that the Lyapunov exponent defining the exponential
evolution of the initial wavefunction with decreasing \newline $n=N,\;N-1,\;N-2,\ldots$ is indeed $-\lambda (h)$ is explicitly verified in Appendix A.  On the other
hand, the Schršdinger equation (\ref{eq.1}-\ref{eq.3}) shows that the rates of hopping from  a given site to its left- and right neighbours, respectively, are
interchanged when $h$ is replaced by $-h$.  This implies that the rates of exponential evolution of the initial wavefunction are also interchanged when $h$ is
replaced by $-h$ i.e. the antisymmetry property \eqref{eq.31} holds.  Finally, we recall that according to the Mott-Borland
[1,2,18] matching condition the symmetrically growing (or decaying) initial wavefunction in the two~directions on the ring,
respectively, lead to proper localized eigenstates at energies corresponding to complex eigenvalues of
(\ref{eq.1}-\ref{eq.3}).\par

The further detailed properties of the IDLL defined for an infinitely large ring ($Nc\rightarrow\infty$) are as follows. For $N|h|c>>1$ \eqref{eq.21} reduces 
to the central limit result

\begin{equation}\label{eq.32}
\lambda=\lambda_\infty\;\text{sign}\;h,
\end{equation}

\noindent where $\lambda_\infty$, which is independent of the sign of h, is given by 

\begin{align}\label{eq.33}
\lambda_\infty &=-\frac{\varepsilon_0^2}{4}\frac{\tilde a}{(1-\tilde a)^2}+c.c.\nonumber\\
&=\frac{\varepsilon_0^2}{4}\biggl(\frac{1}{4-E^2}+c.c.\biggr),
\end{align}

\noindent and defines the IDLL $1/\xi=|\lambda |=|\lambda _\infty |$ at complex energies $E$ of the form (\ref{eq.12}).\par

In the following discussion we regard the Lyapunov exponent $\lambda$ and the IDLL as functions of $ReE$ alone,
which  are obtained by eliminating $ImE$ in favour of $ReE$ using (\ref{eq.13}).  The main properties of the Lyapunov
exponent (\ref{eq.32}-\ref{eq.33}) are:

\newpage
\noindent i.\quad For $h>0$, $\lambda$ is positive in the interval 

\begin{equation}\label{eq.34}
-E_d<ReE<E_d,
\end{equation}
\noindent where
\begin{equation}\label{eq.35}
E_d=\frac{2\cosh\; hc}{\sqrt{1+\tanh^2 hc}}\quad .
\end{equation}

\noindent It possesses a maximum of magnitude 

\begin{equation}\label{eq.36}
\lambda=\frac{\varepsilon_0^2}{8}\frac{1}{1+\sinh^2 hc}
\end{equation}

\noindent at the band centre, $ReE=0$, and vanishes at the values $ReE=\pm E_d$.  The equation (\ref{eq.36}) shows that the
IDLL decreases with increasing $|h|c$ within the domain \linebreak 
$N|h|c\gg 1$.  Finally $\lambda$ is negative in the domains
extending from
$E_d$ to the upper band edge, $2\cosh hc$, and from $-E_d$ to the lower band edge, $-2\cosh hc$, respectively; at the band
edges its value is

\begin{equation}\label{eq.37}
\lambda=-\frac{\varepsilon_0^2}{8\sinh^2 hc}
\end{equation}\\

\noindent ii.\quad For $h<0$, $\lambda$ is negative in the interval (\ref{eq.34}), with a minimum of magnitude (\ref{eq.36}),
and vanishes at $ReE=\pm E_d$. $\lambda$ is positive in the complementary intervalls of (\ref{eq.34}) contained in the real
energy band
$-2\cosh hc\leq ReE\leq 2\cosh hc$ and takes minus the value (\ref{eq.37}) at the band edges.\par

It follows from the above analytical results that the eigenstates of an infinite ring in the complex energy domain (\ref{eq.13}) are
localized, except possible eigenstates with eigenvalues $ReE=\pm E_d$, whose localization lengths would be diverging. 
On the other hand, in the limit of very strong fields ($|h|c>>1$) where the effect of
the disorder becomes negligible, the IDLL \eqref{eq.33} vanishes asymptotically, at
the considered complex energies, as

\begin{equation}\label{eq.38}
\lambda\sim -\frac{\varepsilon^2_0}{2}\cos 2q\;\text{sign}\;h\; e^{-2hc\;\text{sign}\; h}\quad ,
\end{equation}

\noindent so that the wavefunction amplitudes, $\varphi_n\equiv|\varphi_n|\;e^{i\theta_n}=e^{iqn}$ are delocalized in this limit [12].\par

\begin{center}{\bf Application to finite rings}\end{center}

We now turn to the application of the above results in the case of a finite ring of circumference $Nc$.  Here only states whose localization lengths are shorter
than $Nc$ are said to be localized while eigenstates with localization lengths (finite or not) larger than $Nc$ are referred to as delocalized.  The other
important aspect is that the discreteness of the energy spectrum of a finite ring imposes a specific restriction on the validity of the perturbative study of a
weak disorder.  This restriction exists over and above the general strong field requirement which is necessary in order that the disorder may be treated  as a
small  perturbation in both the real and imaginary parts of the energy.  We use the results of our recent calculation of the eigenvalues of (\ref{eq.1}-\ref{eq.3})
\citep{JH2} to explicitate the supplementary condition which must be obeyed by the perturbation parameter of the disorder in a finite ring.  This condition will
allow us to derive an upper bound for the IDLL of the form \eqref{eq.33}.\par

The discrete zeroth order eigenenergies of the ring are given by

\begin{equation}\label{eq.40}
E_0(q_k)=2\cosh hc\cos q_k+2i\sinh hc\sin q_k\;,\; q_k=\frac{2\pi k}{N}\;,\; k=0,\pm 1,\ldots\quad .
\end{equation}

\noindent For weak random site energies with zero mean value and a gaussian correlation, $\langle\varepsilon_i\varepsilon_j\rangle=\varepsilon^2_0\;\delta_{i,j}$,
the averaged second order correction to the energy levels \eqref{eq.40} is given by \citep{JH2}

\begin{equation}\label{eq.41}
\langle E^{(2)}(q_k)\rangle\simeq-\frac{\varepsilon^2_0}{4}\frac{\cot N(q_k-ihc)}{\sin (q_k-ihc)}\quad .
\end{equation}

\noindent The standard heuristic analysis of convergence of non-degenerate Schršdinger perturbation theory \citep{Kramers} now shows that $\langle
E^{(2)}(q_k)\rangle$ must be small compared with the separation $\Delta E_0 (q_k)=E_0(q_{k+1})-E_0(q_k)$ of the unperturbed energy levels \eqref{eq.40} for
convergence to be ensured.  By separating the real and imaginary parts in (\ref{eq.40}-\ref{eq.41}) the convergence conditions of the perturbation theory read

\begin{subequations}
\renewcommand{\theequation}{\theparentequation.\alph{equation}}\label{eq.42}
\begin{align}
|Re\langle E^{(2)}(q_k)\rangle|&<<|\Delta Re E_0(q_k)| \quad ,\\
|Im\langle E^{(2)}(q_k)\rangle|&<<|\Delta Im E_0(q_k)| \quad ,
\end{align}
\end{subequations}

\noindent where

\begin{align}\label{eq.43}
\Delta E_0 (q_k) 
&\simeq\frac{4\pi}{Nc}\left(-\cosh hc\sin q_k+i\sinh hc\cos q_k\right)\quad ,\\
\langle E^{(2)} (q_k)\rangle 
&=\frac{\varepsilon_0^2}{4}\text{sign}\;h\frac{\left(-\cos q_k\sinh hc+i\sin q_k\cosh hc\right)}{\sin^2 q_k+\sinh^2 hc}\quad ,
\end{align}

\noindent for $N|h|c>>1$.\par

It follows from (\ref{eq.43}-41) that the inequalities (\ref{eq.42}.a,b) are not both fulfilled for $q_k=0,\pi$ and for $q_k=\pi/2$ i.e. at the band edges along the
real and complex energy axes, respectively.  For illustration of the condition (\ref{eq.42}.a,b) at typical intermediate \newline $q_k$-values we choose
$q_k=\pi/4$.  For this value the Lyapunov exponent given by \eqref{eq.33} and \eqref{eq.12} takes the form

\begin{equation}\label{eq.44}
\lambda_\infty=\frac{\varepsilon_0^2}{4\cosh^2 (2hc)}\;,\;q_k=\frac{\pi}{4}\quad .
\end{equation}

\noindent Using (41-\ref{eq.44}), the inequality (\ref{eq.42}.a) may be rewritten in terms of the loca\-lization length $\xi=|\lambda_\infty|^{-1}$ in the form

\begin{equation}\label{eq.45}
\xi >>\frac{Nc}{2\pi}\frac{|\tanh hc|\cosh^2(2hc)}{1+2\sinh^2 hc}\;,\;q_k=\frac{\pi}{4}\quad ,
\end{equation}

\noindent while (\ref{eq.42}.b) takes the same form with $|\tanh hc|$ replaced by $|\coth hc|$, thus implying a larger lower bound for $\xi$ than \eqref{eq.45}.  As
mentioned earlier the perturbative treatment of the disorder is generally valid for sufficiently large fields only.  In this context, \eqref{eq.45} shows that the
lower bound of $\xi$, which increases with increasing $h$, is already larger than the ring circumference for fields of relatively moderate strength, of the order of
$hc=1.3$.  This leads to the important conclusion that the complex eigenenergy states in a finite ring are delocalized (and correspond to depinning of the
associated flux line system \citep{Hatano}) in the domain of field strengths where a perturbative treatment of the disorder is valid.  These results provide a
detailed analytical confirmation of the suggestion of Hatano and Nelson that the complex energy eigenstates are delocalized.

\section{CONCLUDING REMARKS}

We have calculated analytically the Lyapunov exponent, $\lambda(h)$, for the eigenstates of a random circular chain in a
strong non-hermitian field $h$, for complex energies, to order $\varepsilon_0^2$ in the disorder.\par

In a finite ring the discrete nature of the energy spectrum entails a specific restriction on the validity of the perturbative treatment of the disorder which
implies the existence of a lower bound for the localization length $\xi=|\lambda(h)|^{-1}$ in \eqref{eq.33}.  This shows that the localization lengths at
intermediate energies within the complex energy band are larger than the circumference of the ring at the strong fields of interest in the perturbation theory.  It
follows that the complex energy eigenstates in a finite ring are delocalized, in agreement with the suggestion of Hatano and Nelson \citep{Hatano}.\par

The analytic argument of HN for showing the existence of a transition from localized to delocalized states in an IVP is based on an imaginary gauge transformation,
$\varphi(x)= \exp (gx/\hbar)\psi(x)$ (with $g/\hbar\equiv h$), which eliminates the IVP, $-ig$, from the Schršdinger equation for their continuous random
hamiltonian for a ring of length $L_x$.  In their analysis HN implicitly replace the twisted boundary condition (which is well-known in the context of real vector
potentials [25,26]) for the transformed wavefunction, namely

\begin{equation}\label{eq.46}
\psi(x+L_x)=e^{-\frac{g L_x}{\hbar}}\psi(x)\quad ,
\end{equation}

\noindent by the periodic boundary confition, $\psi(x+L_x)=\psi(x)$, appropriate for a ring in the absence of $g$.  In general this approximation is valid only for
a very weak IVP such that $\frac{|g|}{\hbar}<<1/L_x$.  HN have obtained localized right and left eigenstates, for weak IVP, from the localized states (with
localization lengths $\xi_n<Nc$) of the ring for $g=0$.  These states are insensitive to the actual boundary condition \eqref{eq.46} if e.g. the localization
length in the case of the right eigenstates, $(1/\xi_n-|g|/\hbar)^{-1}$ \citep{Hatano}, is small compared to $Nc$.  In this case therefore the validity of the
replacement of
\eqref{eq.46} by a periodic boundary condition extends to a range of somewhat larger values of $g$, namely $|g|/\hbar<<1/\xi_n > 1/L_x$.  It remains, however, that
the analytic discussion of HN is restricted to a domain of weak fields for the real eigenvalue localized states.  In contrast the study of energy eigenstates in
Sect.~II and III applies in a domain of large fields, $|h|\geq 1/c$, where the detailed nature of delocalized complex eigenenergy states of a finite ring has not
been studied analytically before.

\newpage
\setcounter{equation}{0}
\renewcommand{\theequation}{A.\arabic{equation}}
\begin{center}{\bf \Large Appendix}\end{center}

Here we calculate the Lyapunov exponent describing the evolution of the wavefunction amplitude at the successive neighbours $N,\;N-1,\;N-2,\ldots,\qquad N-n,\ldots$
of the site~1 along the direction opposite to that considered in Sect.~II.  From successive recursions of (14.a) we obtain

\begin{equation}\label{Aeq.1}
R^{(1)}_{N-n}=\tilde{a}^{-n}R^{(1)}_N+e^{-hc}\sum^n_{m=1}\tilde{a}^{m-n-1}\varepsilon_{N-m}\quad ,\quad n=1,\;2,\ldots,\;N-2\quad ,
\end{equation}

\noindent and from (14.b) and \eqref{Aeq.1} taken at $n=N-1$,

\begin{equation}\label{Aeq.2}
R^{(1)}_N=\frac{e^{-hc}}{1-\tilde{a}^{-N}}\biggl[\tilde{a}^{-2}(\tilde{a}\varepsilon_N+\varepsilon_1)+\sum^{N-2}_{m=1}\tilde{a}^{m-N-1}\varepsilon_{N-m}\biggr]\quad
.
\end{equation}

\noindent Similarly, from (15.a) we find

\begin{equation}\label{Aeq.3}
R^{(2)}_{N-n}=\tilde{a}^{-n}R^{(2)}_N+e^{-iq}\sum^n_{m=1}\tilde{a}^{m-n}R^{(1)2}_{N-m}\quad ,\quad n=1,\;2,\ldots,\;N-2\quad ,
\end{equation}

\noindent and from (15.b) with $R^{(2)}_2$ given by the $n=N-2$ equation of \eqref{Aeq.3},

\begin{multline}\label{Aeq.4}
R^{(2)}_N=\frac{e^{-iq}}{1-\tilde{a}^{-N}}\biggl[(1+\tilde
a)R^{(1)2}_N-e^{iq}(\sqrt{\tilde{a}})^{-1}\varepsilon_N(2\tilde{a}R^{(1)}_N-e^{-hc}\varepsilon_N)\biggr.\\
\biggl.+\sum^{N-2}_{m=1}\tilde{a}^{m-N}R^{(1)2}_{N-m}\biggr]
\end{multline}

\noindent The Lyapunov exponent is defined by (with $\displaystyle{\varphi_{N-n}=\prod^{n-1}_{m=0}R^{-1}_{N-m}Q^{-1}_1\varphi_1}$)

\begin{align}\label{Aeq.5}
\lambda &=\lim_{(n<N)\rightarrow\infty}\frac{1}{nc}\ln|\varphi_{N-n}|\nonumber\\
&= -\lim_{(n\leq N)\rightarrow\infty}\frac{1}{nc}\left(\sum^{n-1}_{p=0}\langle \ln|R_{N-p}|\rangle+\langle\ln|Q_1|\rangle\right)\quad ,
\end{align}

\noindent where $\ln|R_{N-p}|$ is expanded, as in \eqref{eq.19}, to second order in the site energies and the term proportional to $\ln\langle|Q_1|\rangle$
vanishes for $n\rightarrow\infty$.  The disorder averages $\langle R^{(1)2}_{N-p}\rangle,\;\langle R^{(1)2}_N\rangle,\;\langle
R^{(2)}_{N-p}\rangle\;\text{and}\;\langle R^{(2)}_N\rangle$ involved in the expansion of $\langle\ln |R_{N-p}|\rangle$ are computed from the exact solutions
(\ref{Aeq.1}-\ref{Aeq.4}) and are found to coincide with the site-independent values (20.a,b), as expected.  From these results we then obtain

\begin{equation}\label{Aeq.}
\lambda=-\frac{\varepsilon_0^2}{4}\frac{\tilde{a}}{(1-\tilde{a})^2}\left(\frac{1+\tilde{a}^{-N}}{1-\tilde{a}^{-N}}\right)+\text{c.c.}\quad ,
\end{equation}

\noindent which coincides with minus the Lyapunov exponent \eqref{eq.21} along the opposite direction on the ring, as expected.


\newpage
\begin{thebibliography}{99}
\bibliographystyle{unsrt}
\bibitem{Mott} N.F. Mott and W.D. Twose, Adv. Phys. {\bf 10}, 137 (1961).
\bibitem{Borland} R.E. Borland, Proc. Roy. Soc. {\bf  A.274}, 529 (1963).
\bibitem{Thouless} D.J. Thouless, in Ill-Condensed Matter, edited by R. Balian, R. Maynard and G. Toulouse (North Holland
Amsterdam, 1979).
\bibitem{Kappus} M. Kappus and F. Wegner, Z. Phys. {\bf  B45} , 15 (1981).
\bibitem{Lambert} C.J. Lambert, Phys. Rev. {\bf B29 }, 1091 (1984).
\bibitem{Derrida} B. Derrida and E.J. Gardner, J.Phys. (Paris) {\bf  45}, 1283 (1984).
\bibitem{Luck} See also J.M. Luck, Systemes Desordonnes Unidimensionnels (Alea,~Saclay,~1992) and references therein;
J.B.~Pendry, Adv.Phys. {\bf  43}, 461 (1994).
\bibitem{Buttiker} M. BŸttiker, Y. Imry and R. Landauer, Phys.Lett. {\bf 96A }, 365 (1983).
\bibitem{Cheung} H.F. Cheung, Y. Gefen, E.K. Riedel and W.H. Shih, Phys.Rev. {\bf B37 }, 6050 (1988).
\bibitem{Dorokhov} O.N. Dorokhov, Sov.Phys. JETP {\bf  74}, 518 (1992).
\bibitem{Levy} L.P. Levy~{\it et al.} Phys.Rev.Lett. {\bf 64 }, 2074 (1990);
V. Chandrasekhar~{\it et al.}, ibid. {\bf 67}, 3578 (1991); D. Mailly~{\it et al.}, ibid. {\bf 70 }, 2020 (1993).
\bibitem{Hatano} N. Hatano and D.R. Nelson, Phys.Rev.Lett. {\bf 77 }, 570 (1996); ibid. Phys.Rev. {\bf B56}, 8651 (1997).
\bibitem{Efetov} K.B. Efetov, Phys.Rev.Lett. {\bf  79}, 491 (1997); ibid. Phys.Rev. {\bf B56}, 9630 (1997).
\bibitem{Brouwer} P.W. Brouwer, P.G. Silvestrov and C.W.J. Beenakker, Phys.Rev. {\bf  B56}, R4333 (1997).
\bibitem{Goldsheid} I.Y. Goldsheid and B.A. Khoruzhenko,
Phys.Rev.Lett., {\bf 80}, 2897 (1998); C.~Mudry~{\it et al.}, ibid. {\bf  80}, 4257 (1998); N.M. Shnerb and D.R. Nelson, ibid
{\bf 80}, 5172 (1998);J.T. Chalker and B. Mehlig, ibid. {\bf  81}, 3367 (1998); J.S. Caux, ibid. {\bf 81 }, 4196 (1998); P.G.
Silvestrov, ibid. {\bf 82 },3140 (1999); J. Feinberg and A. Zee, Nucl.Phys. {\bf  B 504}, 579 (1997); E. Brezin and A. Zee,
ibid. {\bf  509}, 599 (1998); P.G. Silvestrov, Phys.Rev. {\bf  B 58}, R10111 (1998); N. Hatano and D.R.~Nelson, ibid. {\bf 
58}, 8384 (1998).
\bibitem{Mudry} C. Mudry, P.W. Brouwer and B.I. Halperin, Phys.Rev., {\bf B58}, 13539 (1998).
\bibitem{JH2}  J.~Heinrichs, Phys.~Rev.~{\bf B63}, 165108-1(2001).
\bibitem{Crisanti} A. Crisanti, G. Paladin and A. Vulpiani, Products of Random Matrices in
Statistical Physics (Springer, New~York, 1993).
\bibitem{1}For brievityÕs sake we do not discuss the phase of asymptotic wave functions i.e. the imaginary part of the
characteristic exponent (or complex Lyapunov exponent) which may usually be related to the integrated density of states.
\bibitem{Furstenberg} H. FŸrstenberg, Trans.Ann.Math.Soc., {\bf 108 }, 377 (1963); V.I.~Oseledec, Trans.Moscow
Math.Soc. {\bf 19}, 197 (1968).
\bibitem{Heinrichs} J. Heinrichs, Phys.Rev. {\bf  B51}, 5699 (1995). 
\bibitem{2} We note, however, that the weak disorder expansion is not valid at sites arbitra\-rily far away from the initial
site.   However, it has been shown (for a semi-infinite disordered chain and  $h=0$) that the central limit value for the ILL
is actually attained by iterating to the much shorter distances corresponding to the range of validity of the weak disorder
expansion, provided E differs from the band centre and from the band edges where the ILL is anomalous.  (J. Heinrichs,
Phys.Rev. {\bf B50}, 5295 (1994)).
\bibitem{Ishii} K.ÊIshii, Prog.Theor.Phys. (Supplement), {\bf 53}, {\bf 77 } (1973).
\bibitem{Kramers} See e.g. H.A.~Kramers, Quantum Mechanics (Dover Publications, New York, 1964).
\bibitem{Byers} N.~Byers and C.N.~Yang, Phys.~Rev.~Lett. {\bf 7}, 46 (1961).
\bibitem{Bloch} F.~Bloch, Phys.~Rev. {\bf 137}, A787 (1965); ibid. {\bf 166}, 415 (1968); ibid. {\bf B2}, 109 (1970).
\end{thebibliography}
\end{document}